\documentclass[twocolumn]{aastex7}
\usepackage{hyperref}

\begin{document}

\title{Spectropolarimetric Constraints on the Maunder Minimum Analog HD~166620: \\Evidence for Weakened Magnetic Braking}

\author[0000-0001-5180-2271]{Federica Chiti}
\affiliation{Institute for Astronomy,
University of Hawai‘i at Mānoa,
2680 Woodlawn Dr., Honolulu, HI 96822, USA}
\email{fchiti@hawaii.edu}

\author[0000-0002-4284-8638]{Jennifer L. van Saders}
\affiliation{Institute for Astronomy,
University of Hawai‘i at Mānoa,
2680 Woodlawn Dr., Honolulu, HI 96822, USA}
\email{jlvs@hawaii.edu}

\author[0000-0003-3061-4591]{Oleg Kochukhov}
\affiliation{Department of Physics and Astronomy, Uppsala University, Box 516, SE-75120 Uppsala, Sweden}
\email{oleg.kochukhov@physics.uu.se}

\author[0000-0003-4034-0416]{Travis S. Metcalfe}
\affiliation{Center for Solar-Stellar Connections, WDRC, 9020 Brumm Trail, Golden, CO 80403, USA}
\email{travis@wdrc.org}

\begin{abstract}
We present the first spectropolarimetric time-series analysis of the Maunder Minimum analog HD~166620, using 12 nights of data from CFHT/SPIRou and a single epoch from CFHT/ESPaDOnS. While individual Stokes~$V$ profiles exhibit no significant polarization signatures, we leverage the rotational coverage of the SPIRou dataset to compute a grand average LSD profile. Forward modeling of the cumulative Stokes~$V$ signal, assuming a purely axisymmetric dipole, yields a best-fit dipole field strength of $B_{\rm dip} = 1.10^{+0.95}_{-0.90}$\,G ($3\sigma$). This field strength matches simulations of the solar dipole during the Maunder Minimum phase. Our results are consistent with independent constraints on the dipole field strength from an LBT/PEPSI snapshot and exclude the presence of strong non-axisymmetric fields potentially missed by this single-epoch observation. These findings provide direct empirical evidence that the transition to weakened magnetic braking involves a weakening of the large-scale magnetic field and suggest that HD~166620 represents a state comparable to the Sun near the peak activity of a grand minimum.
\end{abstract}

\keywords{\uat{\uat{Stellar evolution}{1599} --- Stellar magnetic fields}{1610} --- \uat{Stellar Rotation}{1629} --- \uat{Maunder Minimum}{1015} ---\uat{Spectropolarimetry}{1973}}

\section{Introduction} 
Stars like the Sun are born with rapid rotation, but they spin down over their main-sequence lifetimes as magnetized stellar winds carry away angular momentum \citep{weber1967}. This well-established paradigm, first empirically described by the Skumanich relation \citep[$P_{\mathrm{rot}} \propto t^{1/2}$; ][]{skumanich1972}, has long served as the foundation for gyrochronology—the use of rotation periods to determine stellar ages \citep{barnes2007}. The precision of this technique relies on the assumption that magnetic braking operates continuously and predictably throughout a star’s life. 

However, for nearly a decade, we have known that Sun-like stars undergo a rotational crisis in the middle of their main-sequence lives. Upon approaching a critical value of Rossby number (i.e., the ratio between the rotation period, $\mathrm{P_{rot}}$, and the convective overturn timescale, $\tau_{cz}$; Ro $=\mathrm{P_{rot}/\tau_{cz}}$), the efficiency of magnetic braking drastically decreases, causing stars to rotate faster than predicted by standard spin-down. This phenomenon, known as weakened magnetic braking \citep[WMB; ][]{vanSaders2016}, is coincident with the disappearance of Sun-like activity cycles at a similar age and Ro, suggesting a magnetic origin for WMB \citep{metcalfe2017}. The Sun itself resides near this critical threshold, placing our own star in the midst of this magnetic transition \citep{vanSaders2016,metcalfe2017}.
We now know that WMB is not unique to F- and G-type stars \citep{vanSaders2016,metcalfe2019, metcalfe2021,metcalfe2022,metcalfe2023}. It also encompasses K dwarfs and it is triggered as one approaches the same $\mathrm{Ro_{crit}\sim Ro_{\odot}}$ as it is for their hotter F and G counterparts \citep{metcalfe2025a}. This universality confirms the Rossby number as the central variable governing the onset of WMB.

The physical driver of this transition appears to be the collapse of the global stellar dynamo \citep{metcalfe2025b}. This collapse results in a sharp reduction of the large-scale field strength and a shift toward complex, high-order magnetic geometries that are less efficient at braking \citep{reville2015,garraffo2016,garraffo2018}. Characterizing this transition is necessary for calibrating gyrochronology in older populations and understanding the magnetic future of middle-aged stars.

To improve theoretical estimates of braking torques informed by magnetic morphology measurements \citep{Finley2018}, a recent series of studies has targeted stars in the WMB regime using high-resolution spectropolarimetry \citep{metcalfe2019,metcalfe2021,metcalfe2022,metcalfe2023,metcalfe2024a,metcalfe2025a,chiti2025}. Among these targets, in this work we focus on HD~166620. With an age of 9--12 Gyr \citep{baum2022,metcalfe2025a}, HD~166620 is the oldest K dwarf known to reside in the WMB regime. Its Rossby number of $\mathrm{Ro} \approx 1.03\,\mathrm{Ro}_{\odot}$ places it immediately past the critical threshold for the magnetic transition \citep[$\mathrm{Ro_{crit}} \approx 0.9\,\mathrm{Ro}_{\odot}$;][]{saunders2024}. Beyond its rotational state, HD~166620 is unique for its activity state: it is the first identified true analog of the Sun during Maunder Minimum \citep{baum2022,luhn2022}. During this historical epoch (1645--1715), sunspots occurrence decreased substantially \citep{eddy1976}: fewer than 50 were recorded between 1672 and 1699 \citep{spoerer1890}, in stark contrast to the 40,000--50,000 observed in any typical 30-year time frame during the past hundred years \citep{beckman1998}. Mirroring this deep quiescence, recent H~I~Ly$\alpha$ observations of HD~166620 reveal a chromospheric activity level nearly a factor of two lower than the Sun's at solar minimum \citep{wood2026}. As the only star currently known to have entered both this Maunder Minimum phase and the WMB regime like the Sun, HD~166620 offers a unique opportunity for investigating the solar magnetic evolution.

\begin{table*}
    \centering
    \caption{Log of the ESPaDOnS and SPIRou spectropolarimetric observations of HD~166620. The columns provide: (1) the UT date of the observation; (2) instrument used; (3) the rotational phase, calculated relative to the first SPIRou observation and using a rotation period of P$_{\mathrm{rot}}=43$ days \citep{baliunas1996}; (4) the number of independent observations per night; (5) the total observing time per night; (6) the S/N per pixel of the normalized Stokes $I$ spectra averaged across the number of independent observations per night, computed at reference wavelength of 550 nm for ESPaDOnS and 1,750 nm for SPIRou; (7) the polarimetric precision in the coadded Stokes~$V$ LSD profile, computed as the standard deviation of the coadded null LSD profile in the region of the spectral line; (8) the longitudinal magnetic field strength, $\langle B_{\rm z}\rangle$, computed from the coadded LSD Stokes~$V$ profile; and (9) the False Alarm Probability (FAP) of the detection in the coadded Stokes~$V$ LSD profile.}
    \begin{tabular}{ccccccccc}
    \hline
    \hline
        UT Date & Instrument & Phase & Num& $t_{\rm obs}$ [min] & S/N & $\sigma_V \,[\times10^{-5}]$ & $B_{\rm z}\pm\sigma_{B_{\rm z}}$ [G] & FAP $V$  \\
    \hline
    \hline
         2025-04-06& ESPaDOnS & &4 & 55& 610 & $1.38$& $-0.69\pm0.35$& $0.24$\\
         2025-04-11& SPIRou & 0.00 &4 &50 & 895 & $5.31$& $-0.94\pm0.63$ & $0.17$\\
         2025-04-15& SPIRou & 0.09 &4&50 &970 & $7.01$ & $0.61\pm0.78$& $0.25$\\
         2025-04-20& SPIRou & 0.20 &4 & 50&990 & $7.89$& $-0.39\pm0.81$& $0.92$\\
         2025-04-21& SPIRou & 0.23 &4 & 50&660 & $8.06$ & $-0.053\pm0.82$& $0.59$\\
         2025-05-06& SPIRou & 0.57 & 4 &50 &750 & $5.88$& $-0.22\pm0.07$& $0.65$\\
         2025-05-07& SPIRou & 0.59 & 4 & 50&990 & $4.61$& $0.79\pm0.75$& $0.77$\\
         2025-05-10& SPIRou & 0.66 &4 & 50&1000 & $8.10$& $0.30\pm0.85$& $0.97$\\
         2025-05-13& SPIRou & 0.73 & 4 & 50&930 & $5.26$& $-0.50\pm0.78$& $0.56$\\
         2025-05-16& SPIRou & 0.80 & 4 & 50&980 & $8.56$& $-0.72\pm0.73$& $0.28$\\
         2025-05-18& SPIRou & 0.84 & 4 & 50&940 & $6.80$& $-0.39\pm0.81$& $0.37$\\
         2025-06-05& SPIRou & 1.25 & 4 & 50&955 & $5.43$& $-0.46\pm0.14$& $0.54$\\
         2025-06-08& SPIRou & 1.32 & 4 & 50&890 & $4.73$& $-0.71\pm0.14$& $0.92$\\
    \hline
    \end{tabular}
    \label{tab1}
\end{table*}

Recently, \citet{metcalfe2025a} used a single high-resolution spectropolarimetric snapshot from LBT/PEPSI to detect a weak dipole field ($B_{\mathrm{dip}}=-1.10\,$G) in HD~166620 and estimated a wind braking torque consistent with WMB. However, a single snapshot cannot constrain the global magnetic morphology, nor can it rule out the presence of complex, non-axisymmetric fields that may only be visible at specific rotational phases. Determining whether the field is truly globally weak—or simply observed at a local minimum—requires full phase coverage. Thus, in this work, we present the spectropolarimetric time-series of HD~166620, obtained with the SpectroPolarimètre InfraRouge \citep[SPIRou;][Program 25AH95; PI: F. Chiti]{donati2020} near-infrared and the Echelle SpectroPolarimetric Device for the Observation of Stars \citep[ESPaDOnS;][]{donati2003} optical spectropolarimeters at the Canada-France-Hawaii Telescope (CFHT). These observations also allow us to benchmark the capabilities of optical versus near-infrared spectropolarimetry for characterizing the ultra-weak fields of inactive K dwarfs, a regime that challenges the limits of current instrumentation.

\section{Observations}
Observations of HD 166620 were carried out with the spectropolarimeters SPIRou (Section~\ref{sec2.1}) and ESPaDOnS (Section~\ref{sec2.2}) mounted on the 3.6-m Canada-France-Hawaii Telescope (CFHT) on Maunakea.
\subsection{SPIRou Observations}\label{sec2.1}
Observing time was approved for ESPaDOnS but repurposed for SPIRou when scheduling constraints dictated that the instrument would not be mounted for a sufficient amount of time to adequately sample the star's rotation period. We monitored the star on 12 nights between April 11, 2025, and June 08, 2025. SPIRou is a stabilized high-resolution ($R=70,000$) spectropolarimeter that covers the near-infrared domain (\textbf{0.95} to 2.5 $\mu$m) in a single exposure. Our observing strategy was designed to achieve a roughly uniform sampling of the stellar rotation. We adopted a period of $P_{\mathrm{rot}}=43\pm0.5$~d \citep{baliunas1996}, which is consistent with the more recent estimate of $P_{\mathrm{rot}}=45.06\pm4.07$~d from \citet{luhn2022}. A complete log of the observations is provided in Table~\ref{tab1}. The reduced normalized Stokes $I$ spectra of HD~166620 show a typical signal-to-noise ratio (S/N) of 900 per pixel at 1750 nm. 

Each observation consisted of a standard polarimetric sequence of four sub-exposures, where the Fresnel rhomb retarders are rotated to specific azimuths between exposures. This sequence, with individual sub-exposures of $\sim190$ s, allows for the measurement of the unpolarized intensity (Stokes $I$) and circular polarization (Stokes~$V$), while suppressing systematic errors in the polarization spectra to first order \citep{donati1997}. We repeated each sequence four times per observing night, resulting in a total observing time of 50 minutes per night. A diagnostic null spectrum ($N$) was also computed to monitor for instrumental or data reduction artifacts.

The raw observations were reduced using the automated \texttt{APERO} pipeline \citep[v0.7.291;][]{cook2022}, which handles standard calibration and spectral extraction. A significant challenge in this domain is the removal of telluric lines that blanket the near-infrared spectrum. To mitigate this, \texttt{APERO} employs a correction strategy described by \citet{artigau2022} that models atmospheric transmission by fitting the science observations with a combination of synthetic TAPAS templates \citep{bertaux2014} and empirical principal components derived from standard stars monitored throughout SPIRou's operations. Despite this correction, residuals frequently persist in the cores of the deepest absorption bands. To prevent these features from introducing spurious signals into the analysis, we masked regions characterized by strong H$_2$O absorption (typical transmission $<40\%$). Specifically, the following intervals were excluded from the calculation of LSD profiles: [950, 979], [1081, 1169], [1328, 1492], [1784, 2029], and [2380, 2500] nm.

\subsection{ESPaDOnS Observations}\label{sec2.2}
To benchmark the polarimetric precision of SPIRou against optical facilities, which are best suited for studying low-activity, K-type stars, we also acquired spectropolarimetry of HD~166620 on April 4, 2025, using ESPaDOnS at CFHT (Program 25AH12; PI: F. Chiti). ESPaDOnS covers the 370--1050 nm spectral range at a resolving power of $R=65,000$ \citep{donati2003}. We employed the standard polarimetric observing mode, which consists of four sub-exposures taken with different orientations of the Fresnel rhomb retarders, as for SPIRou. To ensure a robust comparison with the near-infrared data, we adopted an exposure time of 205 s, aiming to match the S/N of the SPIRou observations. The data were reduced using the standard \texttt{LIBRE-ESPRIT} pipeline \citep{donati1997}, resulting in a final Stokes $I$ spectrum with an S/N of 610 per pixel at 550 nm.

Similarly to the SPIRou observations, we removed regions with significant telluric absorption prior to analyzing the spectrum: [587, 600], [628, 633], [647, 658], [686, 706], [717, 735], [758, 770], [790, 795], [810, 990], and [840, \textbf{1050}] nm.

\section{Multi-line Polarization Analysis}
To extract the expected weak polarization signatures of HD~166620, we applied Least-Squares Deconvolution \citep[LSD;][]{donati1997,kochukhov2010} through the \texttt{SpecpolFlow} \footnote{\url{https://folsomcp.github.io/specpolFlow/}} open-source pipeline to the reduced spectra from both \texttt{APERO} and \texttt{LIBRE-ESPRIT}. This technique combines the information from thousands of absorption lines available in the echelle orders, resulting in high S/N mean LSD Stokes $I$ and $V$ profiles. LSD assumes that the observed spectrum can be modeled as the convolution of a discrete line mask—computed using the wavelengths, line depths and Landé factors of atomic transitions—with a single, common line profile. Using a weighted linear least-squares technique, the model is fit to the observed spectra to derive a high-precision LSD profile, which acts as a proxy for the mean line profile integrated over the visible stellar disk.

A key requirement of the LSD method is the self-similarity of the spectral lines used for the deconvolution. Consequently, we excluded spectral lines that are too broad and/or deep and deviate significantly from the average line profile. For the SPIRou near-infrared data, we masked the Paschen series (Pa-$\delta$, $\gamma$, $\beta$), Br-$\delta$ [2163 nm], the He triplet at 1083 nm, and strong features from Mg\,I [1183, 1209, 1501, 1714 nm] and Ca\,I [1977 nm]. Similarly, for the optical ESPaDOnS data, we excluded the Ca H \& K lines, the Balmer series (H$\alpha$, $\beta$, $\gamma$, $\delta$), the Mg triplet [516-519 nm], and the Na doublet [589 nm].

\begin{figure}
    \centering
    \includegraphics[width=\linewidth]{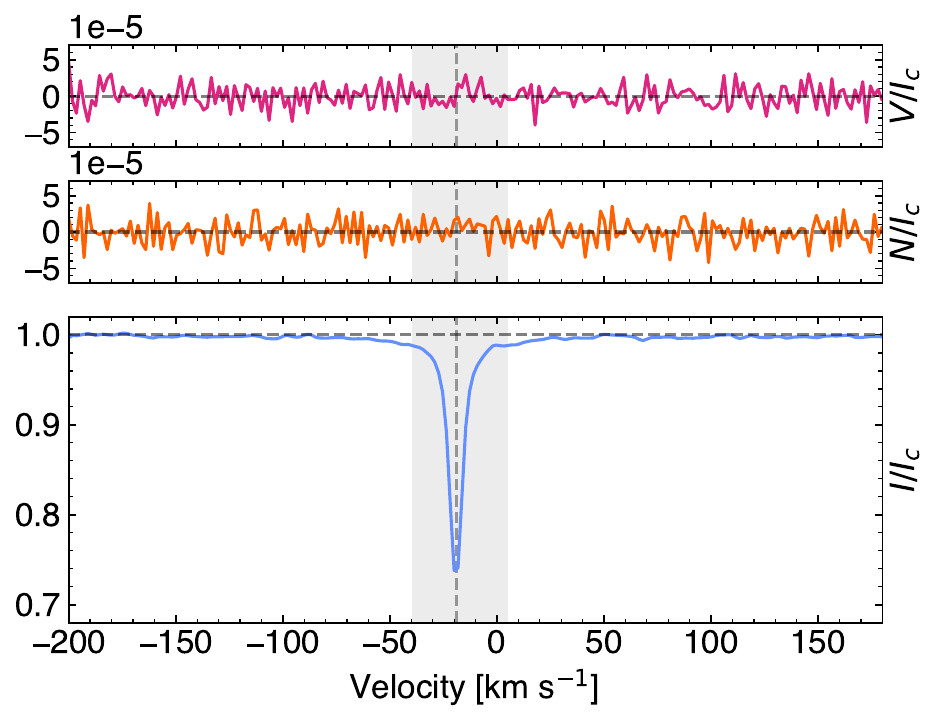}
    \caption{Coadded LSD profiles of HD~166620 derived from ESPaDOnS data showing Stokes~$V$ (top), null profile (middle), and Stokes $I$ (bottom).}
    \label{fig1}
\end{figure}

\begin{figure}
    \centering
    \includegraphics[width=\linewidth]{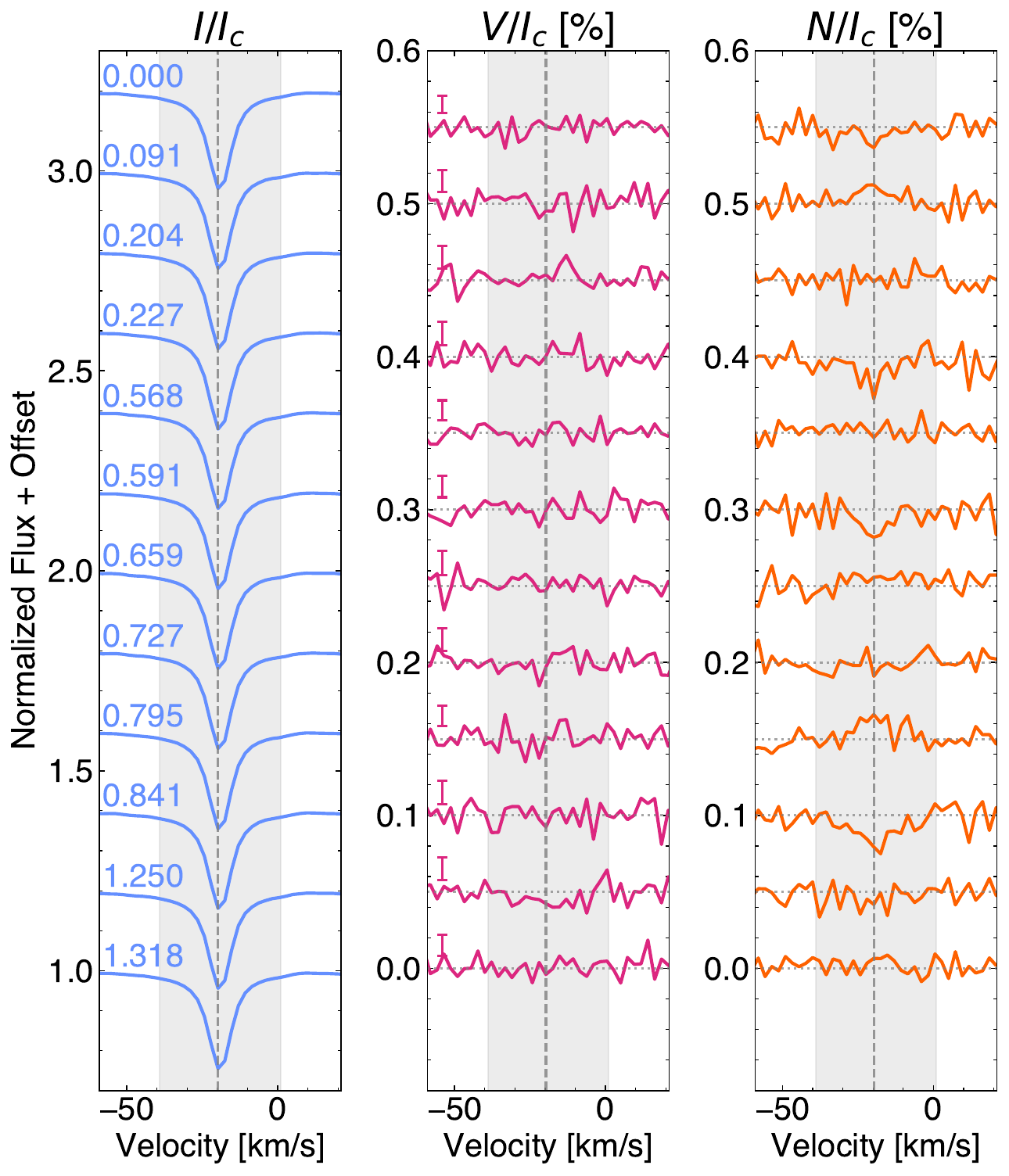}
    \caption{Coadded LSD profiles of HD~166620 derived from SPIRou data. The Stokes $I$ (left), $V$ (middle) and null profile (right) are offset vertically for display purposes. The different rotational phases are shown in the Stokes $I$ panel. Percentage mean errors on the computed LSD  Stokes~$V$ profile are indicated with an error bar in the middle panel. The dashed gray vertical line in all panels marks the radial velocity of the star \citep[$v_{\mathrm{rad}}=-19\,\mathrm{km\,s^{-1}}$;][]{gaia2021}.}
    \label{fig2}
\end{figure}

The line masks were constructed using atomic data extracted from the Vienna Atomic Line Database \citep[VALD;][]{ryabchikova2015}. We selected transitions within the spectral windows of 360-1010 nm (ESPaDOnS) and 950-2600 nm (SPIRou), imposing a minimum central depth threshold of 10\% relative to the continuum. The line lists were computed using an \texttt{ATLAS9} model atmosphere \citep{kurucz1993} with stellar parameters characteristic of HD~166620: $T_{\mathrm{eff}}=5000$ K and $\log g=4.5$ \citep{luhn2022}, and a metallicity of [M/H]$=-0.19$ with a microturbulence of $v_{\mathrm{micro}}=1.0$ km$\,\mathrm{s}^{-1}$ \citep{john2023}. Adopting a VALD line list with [M/H]$=-0.1\pm0.01$, as in \citet{metcalfe2025a}, leads to identical LSD profiles. After filtering out the broad lines listed above and regions of heavy telluric contamination, the final masks comprised 933 lines for the SPIRou data and 8,009 lines for the ESPaDOnS data. We coadded the nightly ESPaDOnS and SPIRou LSD profiles using an inverse-variance ($1/\sigma^2$) weighted average, where $\sigma$ is the uncertainty of the corresponding Stokes parameter (e.g., $\sigma_I$ for Stokes $I$). The coadded LSD profiles are shown in Figures \ref{fig1} and \ref{fig2}, respectively.   

No significant Zeeman signatures were detected in the Stokes~$V$ LSD profiles from either ESPaDOnS or SPIRou. However, the absence of a detection in individual epochs does not preclude the presence of a magnetic field. As demonstrated by \citet{lignieres2009} and \citet{petit2010} in the case of Vega, weak circularly polarized signals that remain buried in the noise of individual LSD Stokes~$V$ profiles can be recovered by summing the signal over all available spectra. \citet{petit2010} further showed that a full time-series of LSD Stokes~$V$ profiles with such sub-noise signatures was sufficient to reconstruct the global magnetic geometry of Vega via Zeeman Doppler Imaging (ZDI). Motivated by this precedent, we proceeded with a forward modeling analysis of the complete time series of SPIRou Stokes $I$ and $V$ profiles to statistically evaluate the evidence for a magnetic field against the null hypothesis.

\section{Forward Modeling of LSD Profiles}

We generated synthetic Stokes $I$ and $V$ LSD profiles using the Zeeman Doppler Imaging (ZDI) code \texttt{InversLSD} \citep{kochukhov2014} and local line profiles computed with the Unno-Rachkovsky analytical solution of the polarized radiative transfer equation \citep{landi2004}. We computed the disk-integrated profile by summing local profiles calculated across a grid of 1,876 independent surface zones \citep{piskunov2002}. As noted by \citet{kochukhov2010}, the normalization of the LSD profiles is somewhat arbitrary, provided that the chosen normalization parameters are applied consistently across the entire analysis. We adopted a mean Landé factor of 1.2, a central wavelength of 1750\,nm and a line depth of $d_0=1.2$ (arbitrary units). The stellar parameters needed for the forward models were chosen from the literature: a projected rotational velocity of $v\,\sin\,i=0.1\,\mathrm{km\,s^{-1}}$ \citep{brewer2016}, a radial velocity of $v_{\mathrm{rad}}=-19\,\mathrm{km\,s^{-1}}$ \citep{gaia2021}, and an inclination of $i=37^{\circ}$ \citep{metcalfe2025a}. We accounted for limb darkening using $H$-band coefficients from a quadratic limb darkening law based on the least-squares method \citet{claret2011}. 

First, we optimized the intrinsic line profile model by adjusting the Doppler width, Voigt damping parameter, and central line depth to achieve the best visual fit to the observed Stokes $I$ LSD profiles. With the Stokes $I$ LSD profile constrained, we proceeded to model the Stokes~$V$ signatures. The global magnetic field geometry in \texttt{InversLSD} is parameterized via a spherical harmonic expansion \citep{kochukhov2014}, which describes the field with three components: radial poloidal (specified by the harmonic coefficient $\alpha_{lm}$, with angular degree, $l$, and azimuthal order of each mode, $m$), horizontal poloidal ($\beta_{lm}$) and horizontal toroidal ($\gamma_{lm}$). The $\alpha_{lm}$ coefficients are related to the dipole magnetic field strength by the following relation
\begin{equation}\label{eq1}
B_{\mathrm{dip}} =\sqrt{\sum_{m=-1,0,+1}^{l=1} \left( \alpha_{l,m} C_{l,m} \right)^2}\end{equation}
where $\alpha_{l,m}$ are the harmonic coefficients for the dipole mode ($l=1$), and the constant $C_{l,m}$ is defined as
\begin{equation}C_{l,m} = \sqrt{\frac{2l+1}{4\pi} \frac{(l-|m|)!}{(l+|m|)!}}\end{equation}
Given the absence of a detectable signal in our observational data, we restricted our forward modeling to the simplest possible configuration: a pure, axisymmetric dipole (with purely poloidal components $\beta_{lm}=\alpha_{lm}$ and $\gamma_{lm}=0$; $l_{\mathrm{max}}=1$, $\alpha_{1,-1}=\alpha_{1,+1}=0$ and $\alpha_{1,0}\neq0$). For this specific test, Equation~\ref{eq1} becomes $B_{\mathrm{dip}} = \alpha_{1,0} \sqrt{\frac{3}{4\pi}}$. We generated a grid of synthetic Stokes~$V$ profiles for a purely axisymmetric dipole with $\alpha_{1,0}$ ranging from $\pm10^{-5}$\,G to $\pm100$\,G.

To assess the significance of magnetic field detection, we employed a chi-square minimization approach that compares our axisymmetric dipole models to the observed Stokes~$V$ profiles. For each trial $\alpha_{1,0}$ value, we computed the total $\chi^2$ statistic summed across all 12 observations

\begin{equation}
    \chi^2 = \sum_{j=1}^{12}\sum_{i}^{N} \left( \frac{V_{\mathrm{obs},ij}-V_{\mathrm{mod},ij}}{\sigma_{V_{\mathrm{obs},ij}}} \right)^2
\end{equation}
where $V_{\mathrm{obs},ij}$ and $V_{\mathrm{mod},ij}$ represent the observed and modeled Stokes $V$ values, respectively, at the $i^{\mathrm{th}}$ velocity point of the $j^{\mathrm{th}}$ observing night. The term $\sigma_{V_{\mathrm{obs},ij}}$ denotes the associated observational uncertainty, and $N$ is the total number of velocity points in each LSD profile. We calculated the difference $\Delta\chi^2=\chi^2-\chi^2_{\mathrm{min}}$ for each trial $\alpha_{1,0}$ value, where $\chi^2_{\mathrm{min}}$ is the minimum chi square across all trial $\alpha_{1,0}$ values. This $\Delta\chi^2$ profile reveals both the detection significance (by comparing the zero-field model to the best fit) and the confidence intervals on $\alpha_{1,0}$ (through the standard thresholds of $\Delta\chi^2= 1.0,\,4.0,\,9.0$ for $1\sigma,\,2\sigma,\,3\sigma$ confidence levels, respectively).

To validate that any detected signal originates from genuine Zeeman signatures rather than instrumental artifacts or systematic effects, we performed identical chi-square analysis on the null profiles. The null profiles should exhibit no Zeeman signature and remain consistent with pure noise. Any systematic instrumental effects would manifest equally in both Stokes~$V$ and null profiles and produce similar $\chi^2$ minima at the same field strengths. By contrast, a genuine magnetic signal produces distinctly different $\Delta\chi^2$ profiles: the Stokes~$V$ curve should minimize at the distinct non-zero value of the detected stellar magnetic field, while the null profile curve should remain consistent with pure noise (minimizing at $\alpha_{1,0}=0$\,G) and rise sharply at non-zero field strengths, where the model predicts a signal that does not exist in the null diagnostic profile.
\begin{figure}
    \centering
    \includegraphics[width=\linewidth]{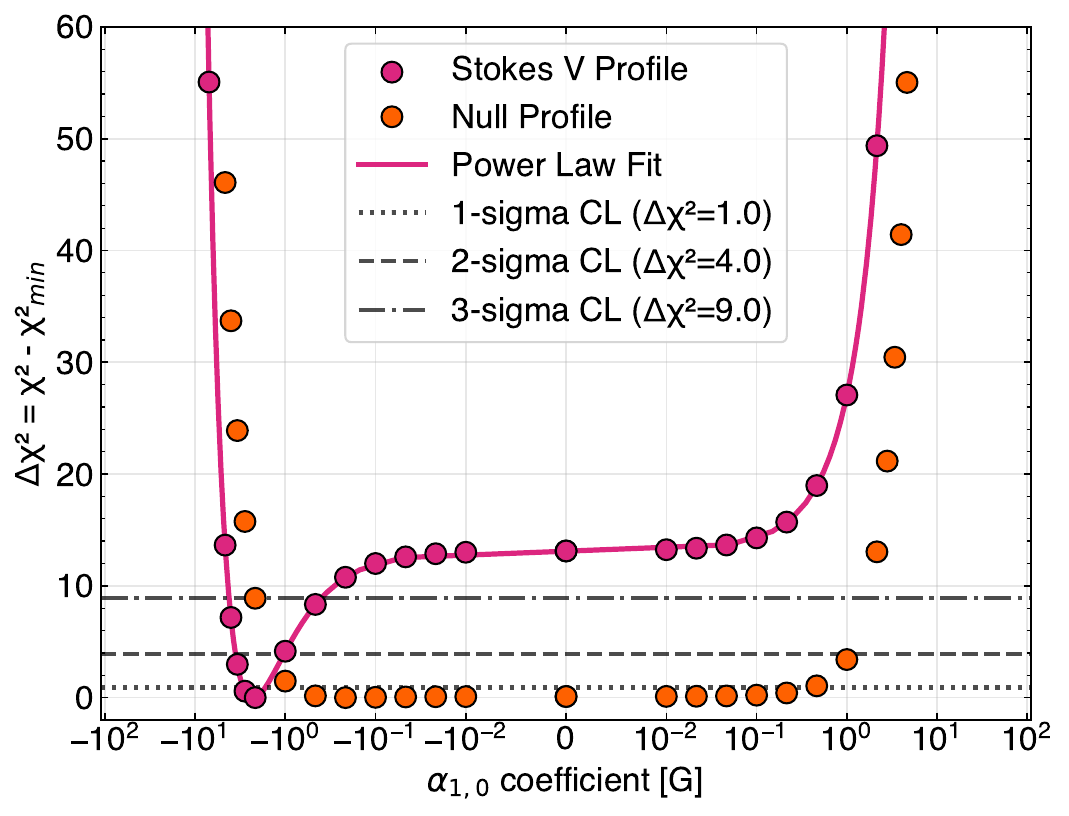}
    \caption{Chi-square difference ($\Delta\chi^2 = \chi^2-\chi^2_{\mathrm{min}}$) as a function of spherical harmonic coefficient $\alpha_{1,0}$ (plotted on a symmetrical logarithmic $x-$scale) for axisymmetric dipole models compared to observed Stokes~$V$ and null  LSD profiles. Horizontal lines indicate confidence level (CL) thresholds. A power law fit is shown for $\Delta\chi^2$ of Stokes~$V$.}
    \label{fig3}
\end{figure}

Figure~\ref{fig3} displays the $\Delta\chi^2$ landscapes for both the Stokes~$V$ and null profiles when compared to the tested purely axisymmetric dipole model. The null profile curve shows no minimum; it remains flat near zero for $\alpha_{1,0}$ values between $\pm 1$\,G and rises sharply outside this range ($\Delta\chi^2 \gg 10$ for $|\alpha_{1,0}| > 1$\,G), confirming the expected consistency with the no-field hypothesis for the null profiles. In contrast, the Stokes $V$ curve shows a well-defined minimum. To precisely determine the $\alpha_{1,0}$ value at this minimum, we fitted a power law of the form $y = a |x - x_0|^p + b$ to the Stokes~$V$ $\Delta\chi^2$ distribution. 
In this functional form, the absolute value term naturally accommodates the V-shaped profile observed in the data, and the power law exponent $p$ provides flexibility to capture the asymmetric steepness of the chi-square surface on either side of the minimum. We found best-fit parameters of $a=2.5$, $b=-0.1$, $x_0=-2.3$, and $p=2.0$, yielding a best-fit $\alpha_{1,0}=-2.25$\,G. The intersections of this power-law fit with the standard confidence level thresholds ($\Delta\chi^2 = 1.0$, 4.0, and 9.0) define the confidence intervals for the $\alpha_{1,0}$ coefficient, as reported in Table~\ref{tab2}. At the Stokes~$V$ best-fit $\alpha_{1,0}=-2.25$\,G, the null profile yields $\Delta\chi^2 \approx 10$, while the Stokes~$V$ profile shows $\Delta\chi^2 \approx 0$, demonstrating that the Zeeman signature is present exclusively in Stokes~$V$. This different behavior confirms that the polarization signal observed in Stokes~$V$ is statistically distinct from instrumental noise and is of stellar origin.

\begin{table}[]
    \centering
    \begin{tabular}{ccc}
    \hline
    \hline
         Best-fit [G]& Confidence Interval [G]& Confidence Level \\
    \hline
    \hline    
         $-2.25$ & $-2.90, -1.60$ & 68.3\% (1$\sigma$)\\
          & $-3.60, -1.00$ & 95.5\% (2$\sigma$)\\
          & $-4.20, -0.40$ & 99.7\% (3$\sigma$)\\
    \hline
    \end{tabular}
    \caption{Confidence limits on the best-fit $\alpha_{1,0}$ harmonic coefficient.}
    \label{tab2}
\end{table}

\begin{figure}
    \centering
    \includegraphics[width=\linewidth]{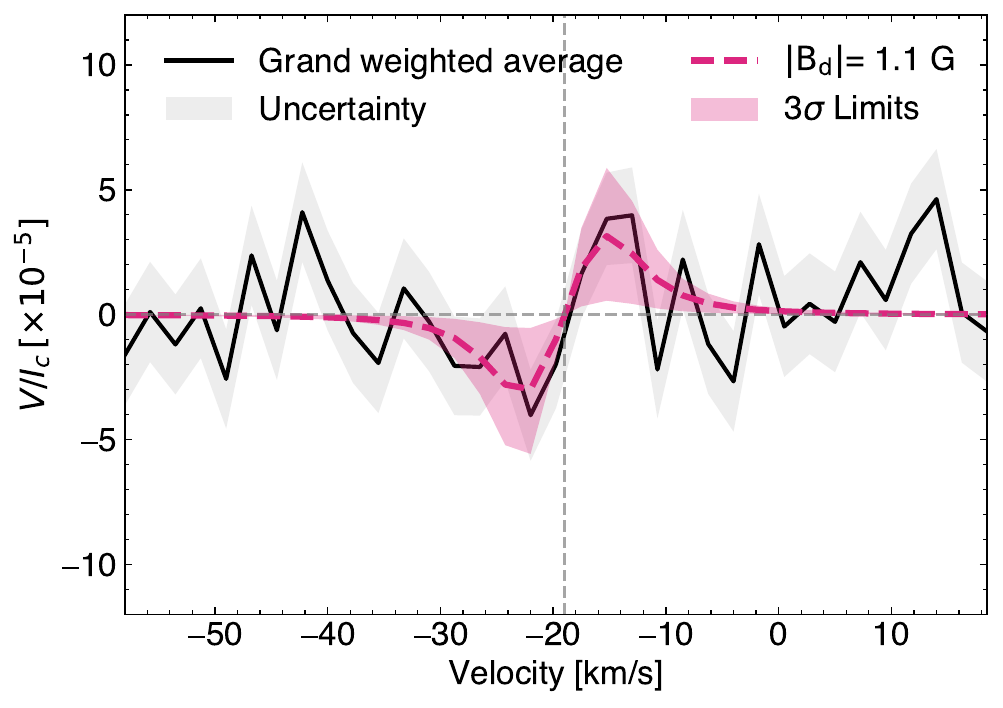}
    \caption{Grand average with uncertainties of the SPIRou Stokes~$V$ polarization profiles for HD~166620. The dashed magenta line is the best-fit axisymmetric model profile of a dipole with $B_{\mathrm{dip}}=1.10\,$G and fixed inclination.}
    \label{fig4}
\end{figure}

We converted the best-fit $\alpha_{1,0}$ coefficient into a best-fit dipole field strength $B_{\mathrm{dip}}$ using Equation~\ref{eq1}, which yielded $B_{\mathrm{dip}} = 1.10$\,G. The corresponding $3\sigma$ confidence limits restrict the field strength of the axisymmetric dipole to the range $0.20\,\mathrm{G} \le B_{\mathrm{dip}} \le 2.05\,\mathrm{G}$. The model reproduces the observations with residual standard deviations of $\sigma_I = 4.0\times10^{-2}$ and $\sigma_V=1.2\times10^{-4}$ for the intensity and polarization profiles, respectively. Figure~\ref{fig4} shows the observed grand average profile alongside the synthetic profile of an axisymmetric dipole with the best-fit and $3\sigma$ boundary field strengths.

\section{Discussion \& Conclusions}
The rotational and magnetic evolution of Sun-like stars is observed to undergo a fundamental transition as stars reach the middle of their main-sequence lifetime. When the Rossby number approaches a critical threshold, stars enter the WMB regime, where their large-scale fields weaken significantly, diminishing the efficiency of magnetic braking. In this paper, we present the first spectropolarimetric time series of HD~166620, the only known star to have entered both a Maunder Minimum phase and the WMB regime, like the Sun. 

We monitored HD~166620 with SPIRou on 12 nights over two months to sample its 43-day rotation period. While individual LSD Stokes~$V$ profiles yielded no detections, a faint signal is expected given the 20 ppm amplitude observed by \citet{metcalfe2025a} with LBT/PEPSI. To search for this, we computed a grand average profile from all 48 observations and performed a forward-modeling analysis using pure axisymmetric dipole models. The $\Delta\chi^2$ landscape reveals a distinct minimum at $\alpha_{1,0} = -2.25$\,G, with a $3\sigma$ confidence interval spanning $-4.2$ to $-0.4$\,G (Figure \ref{fig3}). This corresponds to a best-fit dipole field strength of $B_{\mathrm{dip}} = 1.10$\,G, constrained to the range $0.20\,\mathrm{G} \le B_{\mathrm{dip}} \le 2.05\,\mathrm{G}$ at the 3$\sigma$ level. In contrast, the null profile statistic minimizes near zero field and rises sharply for $|\alpha_{1,0}| > 1$\,G, confirming the stellar origin of the Stokes~$V$ signal.

We estimate the magnetic field strength expected from standard braking by assuming it scales with photospheric pressure ($P_{\mathrm{phot}}$) and Rossby number as $B/B_{\odot}=(P_\mathrm{phot}/P_\mathrm{{phot,\odot}})^{1/2}\,\mathrm{(Ro/Ro_{\odot})}^{-1}$ \citep{vansaders2013}. Using a standard braking law \citep{vansaders2013} for a solar metallicity, $0.8\,\mathrm{M_{\odot}}$ track with the setup described by \cite{chiti2024}, we find $\mathrm{Ro/Ro}_{\odot} = 0.997$ and $B/B_{\odot} = 1.16$ for HD~166620. Adopting the solar dipole field strength of $B_{\mathrm{dip},\odot}=1.54\pm0.66\,$G \citep{Finley2018}, this yields a predicted dipole field of $1.78\,$G. This value is consistent within $3\sigma$ with the dipole strength we inferred from forward modeling. However, HD~166620 has a Rossby number of $\mathrm{Ro} = 1.033^{+0.071}_{-0.065}\,\mathrm{Ro}_{\odot}$, that places it near the critical threshold for the dynamo transition \citep[$\mathrm{Ro_{crit}} \approx 0.9\,\mathrm{Ro}_{\odot}$;][]{saunders2024} and within the WMB regime. Furthermore, based on torque estimates, \citet{metcalfe2025b} demonstrated that, similar to the Sun, this target sits on the cusp of the dramatic drop in observed versus predicted torque experienced by stars in the WMB phase. Together, these results provide evidence that HD~166620 is in an evolutionary stage very similar to that of our Sun.  

As the first true Maunder Minimum analog, HD~166620 provides a unique opportunity to compare stellar magnetic properties with the Sun during its own grand minimum phase. While direct measurements of the Sun's dipole field strength during the Maunder Minimum (1645-1715) do not exist, \cite{wang2003} employed a flux transport model to simulate the evolution of the Sun's magnetic field under Maunder Minimum conditions. They found that the Sun's axial dipole component peaked at an amplitude of approximately 0.5~G, substantially weaker than the $\sim 4\,$G observed during solar cycle 21. Combining their simulated axial and equatorial dipole components yields a total dipole field strength ranging from approximately 0.3-0.4 G (minimum) to 1.3-1.5 G (maximum), with a typical value of 0.6-0.8 G during the Maunder Minimum (see their Figure 5 for more details). Our inferred dipole field strength of 1.10 G for HD~166620 falls within the upper range of the Sun's Maunder Minimum field strength, suggesting that HD~166620 may represent a state comparable to the Sun near the peak activity of its grand minimum phase. 

The resemblance between HD 166620 and the Sun extends beyond their grand minimum states. Recent \textit{XMM-Newton} observations reveal that the corona of HD 166620 is consistent with the emission levels of the solar background corona \citep{bennedik2026}---the baseline X-ray flux state the Sun occupies during the activity minima of its 11-year cycle \citep{peres2000}. During these solar minima, the Sun's large-scale magnetic field is dominated by an axisymmetric dipole \citep{derosa2012,zieger2019}. In this work, we showed that the weak polarization signature of HD 166620 is likewise consistent with forward modeling of a purely axisymmetric dipole, reinforcing the hypothesis that HD 166620 is a close analog of the modern quiet Sun in terms of its large-scale magnetic field as well.

Lastly, our estimate of $B_{\mathrm{dip}}$ is in remarkable agreement with the $B_{\mathrm{dip}} = 1.10^{+0.42}_{-0.40}$\,G measured by \citet{metcalfe2025a} from a single LBT/PEPSI snapshot. While the lower S/N of our SPIRou observations may be preventing us from resolving low amplitude rotational modulation in the magnetic field strength, if the dominant large-scale field component is truly a stable, axisymmetric dipole, the agreement between the SPIRou time-series and the PEPSI snapshot is expected. Due to limited phase coverage, \citet{metcalfe2025a} adopted the conservative assumption of a dominant axisymmetric dipole to estimate the wind braking torque of this star. Our spectropolarimetric time series validates this assumption, ruling out the presence of strong non-axisymmetric components that a single snapshot might have missed. 

The consistency between the dipole field strengths inferred from a single spectropolarimetric snapshot and a time-series in HD~166620 parallels recent results for another star in the WMB regime, $\tau$\,Ceti. Using an ESPaDOnS time series of $\tau$\,Ceti, \cite{chiti2025} confirmed the dipole field strength previously hinted at by a single PEPSI snapshot. Taken together, these results suggest that the strength of the large-scale magnetic fields of these old, quiet Sun-like stars is remarkably stable. Thus, sparse sampling---or potentially even single snapshots---may be sufficient to reliably constrain the large-scale field strength of stars with long rotation periods and flat activity like HD~166620.

However, constraining the magnetic morphology of such inactive stars represents a significant observational challenge. The Stokes~$V$ amplitude in HD~166620's PEPSI snapshot was only 20 ppm \citep{metcalfe2025a}. Resolving such a minute signal is difficult for SPIRou ($R=70,000$), which has roughly half the resolving power of PEPSI ($R=130,000$). Furthermore, while SPIRou has proven highly effective for characterizing magnetic fields in M dwarfs \citep[e.g.,][]{donati2023,fouque2023,lehmann2024}, its optimization for cooler targets makes it less ideal for early K-dwarfs. Although \citet{moutou2020} found comparable performance between SPIRou and ESPaDOnS for \textit{active} K-dwarfs, our analysis highlights the difficulty of observing an \textit{inactive} target like HD~166620. While SPIRou benefits from the stronger Zeeman effect in the infrared ($V \propto \lambda$), this theoretical advantage is outweighed by an order of magnitude larger number of spectral lines available for LSD extraction in the optical for this target. We find that the polarimetric precision in a single ESPaDOnS snapshot ($\sigma_V \approx 1.38 \times 10^{-5}$) is nearly a factor of 4 better than the average precision achieved with SPIRou ($\sigma_V \approx 5 \times 10^{-5}$). Even after accounting for the favorable infrared Zeeman scaling, the line-count advantage of ESPaDOnS results in a higher sensitivity to ultra-weak magnetic fields, a capability recently demonstrated by the detection of a sub-Gauss field ($B_{\mathrm{dip}}=0.31$\,G) on the G8V star $\tau$ Ceti using ESPaDOnS \citep{chiti2025}. 

An additional barrier to characterizing the magnetic field morphologies of old Sun-like stars in the WMB regime is practical: monitoring stars with long rotation periods requires instruments to be mounted on the telescope for extended durations. In recent years, the high demand for SPIRou has limited the availability of ESPaDOnS, making long-baseline optical campaigns difficult. A potential solution may arrive in mid-2026 with the installation of VISION (a.k.a. Wenaokeao), a new device at CFHT that co-mounts the SPIRou and ESPaDOnS Cassegrain units \citep{moutou2025}. This will allow the instruments to be used consecutively or simultaneously, significantly increasing their availability for time-domain studies. By combining the high optical precision of ESPaDOnS with the infrared coverage of SPIRou, VISION will provide the sensitivity and scheduling flexibility required to map the subtle, evolving magnetic fields of stars entering WMB.

\vspace{0.5cm}

\noindent F.C. thanks Luc Arnold for helpful discussion on the SPIRou data reduction products. F.C and J.v.S. acknowledge support from NSF grant AST-2205888. O.K. acknowledges funding by the Swedish Research Council (grant agreement 2023-03667) and the Swedish National Space Agency. T.S.M.\ acknowledges support from NSF grants AST-2205919 and AST-2507890. This work has made use of the VALD database, operated at Uppsala University and the University of Montpellier. We thank Dr. T. Ryabchikova for her invaluable work with compiling and assessing atomic data for the VALD database. Based on observations obtained at the Canada-France-Hawaii Telescope (CFHT) which is operated by the National Research Council (NRC) of Canada, the Institut National des Sciences de l'Univers of the Centre National de la Recherche Scientifique (CNRS) of France, and the University of Hawaii. The observations at the CFHT were performed by the QSO team with care and respect from the summit of Maunakea which is a significant cultural and historic site.

\facilities{CFHT (SPIRou, ESPaDOnS)}
\software{\texttt{SpecpolFlow} \citep{folsom2025}; \texttt{InversLSD} \citep{kochukhov2014}
          }

\bibliography{main}{}
\bibliographystyle{aasjournal-compact}

\end{document}